\documentclass[journal,draftcls,onecolumn,11pt]{article}
   \usepackage{amssymb}
  \usepackage{graphicx}
  \usepackage{ifpdf}
  \pdfoutput=1
\begin{document}

\begin{center}
{\LARGE
On the role of symmetry in solving maximum lifetime problem in two-dimensional sensor networks
}
\end{center}

\begin{center}
Z. Lipi{\'n}ski\\
Institute of Mathematics and Informatics, University of Opole
\end{center}

\begin{abstract}
We analyze a continuous and discrete symmetries of the maximum lifetime problem in two dimensional
sensor networks.
We show, how a symmetry of the network and invariance of the problem under a given transformation
group $G$ can be utilized to simplify its solution.
We prove, that for a $G$-invariant maximum lifetime problem there exists a $G$-invariant solution.
Constraints which follow from the $G$-invariance allow to
reduce the problem and its solution to the subset, an optimal fundamental region of the sensor network.
We analyze in detail solutions of the maximum network lifetime problem
invariant under a group of isometry transformations of a two dimensional Euclidean plane.
\end{abstract}
\section{Introduction}
Let us denote by $S_{N}^{K}$ a sensor network build of $N$ sensors and $K$ data collectors.
We split the set $S_{N}^{K}$ into two subsets,
the set of data collectors $C_{K}$ and the set of sensors $S_{N}$, such that $C_{K}\cup S_{N}$.
Elements $p_i$ of the network $S_{N}^{K}$ we identify with points $p_i=(p^1_i,p^2_i)$ of a two dimensional plane $R^2$,
where $p_i\in C_{K}$ for $i\in [1,K]$ and $p_i\in S_{N}$ for $i\in [K+1,K+N]$.
Each sensor $p_i\in S_{N}^{}$ generates periodically the amount $Q_i$ of data
and sends it, possibly via other sensors, to the data collectors.
The data transmission cost energy matrix $E_{i,j}$
defines the energy required to send one unit of data between two elements
$p_{i}$, $p_{j}$ of the network $S_{N}^{K}$.
The energy consumed by the $p_i$ sensor to send all of its data
in a one cycle of the network lifetime is given by the formula
\begin{equation} \label{SensorEnergy}
E_{i}(q,\bar{p})=\sum_{j=1, j \neq i}^{K+N} q_{i,j}(\bar{p}) E_{i,j}(\bar{p}),
\end{equation}
where $q_{i,j}(\bar{p})$ is the amount of data send by the $p_{i}$ sensor to the $p_{j}$
element of the network $S_{N}^{K}$ and $\bar{p}=(p_1,....,p_{K+N})$.
By definition the data collectors $p_i$ are elements of the network $S_{N}^{K}$
which do not send any data, i.e.,
$$\forall p_{i}\in C_{K}^{} \;\forall p_{j}\in S_{N}^{K} \;\;q_{i,j}(\bar{p})=0.$$ 
From the above assumption it follows that for the data collectors the energy given by
(\ref{SensorEnergy})
is equal to zero, $E_{i}(q,\bar{p})=0$, $i\in [1,K]$.
Because the sensors have limited resources of energy, to extend the network lifetime
we need to find such graph for the data transmitted in the network
that the energy consumed by the most overloaded sensor would be minimal.
Namely, if we assume that all sensors have the same initial energy $E_{0}$ and
for a given data transmission graph the most overloaded sensor 
consumes in a one cycle $E_{i}^{\max}$ of energy,
then $[E_{0} /E_{i}^{\max}]$ is the number of cycles until this sensor runs out of energy.
We define the network lifetime as a number of cycles the data can be transmitted in the network until the first sensor
runs out of energy, \cite{Chang}, \cite{Giridhar}.
In this paper we will not discuss a particular solutions
of the maximum network lifetime problem
and for our purposes we consider a simplified version of it.
We assume, that the only initial parameters which define the problem
are data transmission cost energy matrix $E_{i,j}(\bar{p})$ and the amount of data generated by each node $Q_{i}(\bar{p})$.
For example, we assume that the initial energy of each sensor and the
capacity of transmission channels between elements of the network $S_{N}^{K}$ are sufficiently large
that at least one solution of the problem exists.
We also assume, that the sensors can send their data to any element of the network $S_{N}^{K}$
and the data collectors can receive any amount of data without costs.
Under the above assumptions the maximum lifetime problem for a sensor network $S_{N}^{K}$ can be written in the form
\begin{equation} \label{MinimaxDefMatrix}\left\{   \begin{array}{l}
\min_{q} \max_{E_{i}} \{ E_{i}(q,\bar{p}) \}_{i \in [1,N+K]},  \\
E_{i}(q,\bar{p}) = (q E^{T}(\bar{p}))_{i,i}, \\
(q - q^{T}) v^{}_{0} - Q^{}(\bar{p})=0,\\
E_{i,j}(\bar{p}) \geq 0, Q_{i}(\bar{p}) \geq 0, q_{i,j}(\bar{p}) \geq 0,\\
\end{array} \right.
\end{equation}
where the vector $v_{0}$ has the form $(\underbrace{-1,...,-1}_{K},\underbrace{1,...,1}_{N})$,
$Q^{}_{}(\bar{p}) = (q_1^{(c)},...,q_K^{(c)},Q_{K+1}(\bar{p}),...,Q_{K+N}(\bar{p}))$
and $E^{T}$ is a transposition of the data transmission cost energy matrix $E$.
The undetermined numbers $q_k^{(c)}$ satisfy
$q_k^{(c)} = \sum_{i}^{} q_{i,k}(\bar{p})$, $\sum_{k=1}^{K} q_k^{(c)}=\sum_{n=1}^{N} Q_{K+n}(\bar{p}).$
The first formula in (\ref{MinimaxDefMatrix}) means, that we minimize the objective function
\begin{equation} \label{ObjFunction}
f(q,\bar{p})=\max_{E_{i}} \{ E_{i}(q, \bar{p}) \}_{i\in [1,K+N] }
\end{equation}
of the maximum network lifetime problem with respect to the $q_{i,j}$ variables.
The second formula in (\ref{MinimaxDefMatrix}) defines the energy consumed by each sensor
to send all of its data in a one cycle of the network lifetime.
This equation is a matrix form of (\ref{SensorEnergy}).
The third formula in (\ref{MinimaxDefMatrix}) is a data transmission flow conservation constraint,
which states that the amount of data $Q_{i}(\bar{p})$ generated by the $p_i$ sensor and the amount of data
received from other sensors $\sum_{j} q_{j,i}$
must be equal to the amount $\sum_{j} q_{i,j}$ of data the $p_i$ sensor can send.
The notation in (\ref{MinimaxDefMatrix}) indicates, that in general
the functions $E_{i,j}(\bar{p})$, $Q_i(\bar{p})$ and the solutions $q_{i,j}(\bar{p})$
may depend not only on the coordinates of the points $p_i$ and $p_j$
but also on the coordinates of other elements of the network $S_{N}^{K}$ and thus the functions
may have a non-local character.

In \cite{Woo} there were identified five power-aware metrics for data transmission in mobile ad-hoc networks,
which can be used to define a network lifetime.
The above definition of network lifetime problem is equivalent to the minimization of the 'maximum node cost',
the fifth metric defined in \cite{Woo}.

In the paper we  discuss two types of symmetries of the maximum network lifetime problem
and impact of these symmetries on the solution of (\ref{MinimaxDefMatrix}).
We show, how a symmetry of the problem (\ref{MinimaxDefMatrix})
and a symmetry of the set $S_{N}^{K}$ can be used to simplify
the solution of (\ref{MinimaxDefMatrix}) and reduce it to some subset of $S_{N}^{K}$.
The first type of symmetry is related to an invariance of the problem (\ref{MinimaxDefMatrix})
under a continuous group of transformation $G$ of the two dimensional plane $R^2$ onto itself
in which the sensor network $S^{N}_{K}$ is embedded.
Under a transformation $g \in G$ the elements $p_i$ of the network $S_{N}^{K}$
are moved to another location $g(p)=p'\in S_{N}^{'K}$ of the plane $R^2$.
We assume, that the numbers of sensors $N$ and data collectors $K$ under these transformation remain unchanged.
If the equations (\ref{MinimaxDefMatrix}) are invariant under transformations group $G$,
then there arises a question whether their solutions $q_{i,j}(\bar{p})$ for $S^{N}_{K}$
and $q'_{i,j}(\bar{p'})$ for $S^{'N}_{K}$ coincide.
We show in Sections 3, that it is indeed the case.
We call this type of symmetry the space symmetry,
because it exhibits the global properties of the functions $E_{i,j}(\bar{p})$,
$Q_{i}(\bar{p})$, $q_{i,j}(\bar{p})$ and the whole problem in $R^2$.
For example, if the matrix elements $E_{i,j}(\bar{p})$ are functions
of the Euclidean distance between points $p_i$ and $p_j$ of $S^{N}_{K}$ network,
for simplicity we assume that $Q_{i}(\bar{p})$ are constant functions,
then the problem (\ref{MinimaxDefMatrix}) is invariant under group
of isometries of the Euclidean plane $R^2$, \cite{Martin}.
Invariance of $E_{i,j}(\bar{p})$ under isometry transformation means
that the cost of data transmission between two elements of $S_{N}^{K}$
does not depend neither on the direction of the data transmission nor location of the network $S_{N}^{K}$ in $R^2$.
This property is called an isotropy property of the problem (\ref{MinimaxDefMatrix}) in $R^2$.
The second type of symmetry is related to an invariance of (\ref{MinimaxDefMatrix})
under transformations group of the finite set $S_{N}^{K}$.
In this paper we consider a bijective transformations of the set $S_{N}^{K}$ onto itself.
By definition, such transformations $g$ are permutations, i.e.  $g(p_i) = p_{g(i)}$,
and form a subgroup of symmetric group $\Pi_{}$ of the set $S_{N}^{K}$, \cite{Grove}.
Because we do not want to mix the sensors and data collectors we assume that
the group $G$ acts separately on the sets $C_{K}$ and $S_{N}$, which means that
\begin{equation} \label{SymmetryGroupForSN}
  G \subseteq \Pi(C_{K}) \oplus \Pi(S_{N}).
\end{equation}
Thus, in the paper we consider transformation groups $G$ of the network $S_{N}^{K}$ which are subgroups of
the symmetric group $\Pi(C_{K}) \oplus \Pi(S_{N})$.
In general, symmetries of the functions $E_{i,j}(\bar{p})$ and $Q_{i}(\bar{p})$ and the whole problem
(\ref{MinimaxDefMatrix}) do not have to be related with the shape of the set $S_{N}^{K}$.
We can establish the relation by requiring that,
for a given symmetry group $G$ of the set $S_{N}^{K}$,
we will consider only a $G$-invariant functions $E_{i,j}(\bar{p})$, $Q_{i}(\bar{p})$ and search
for a $G$-invariant solution of (\ref{MinimaxDefMatrix}).
If this is the case, the problem (\ref{MinimaxDefMatrix}) we call
a problem with an internal symmetry group, because it is related to the shape of the set $S_{N}^{K}$.

\section{Related work}
Symmetries quite naturally appeared in a several well known optimization and combinatorial problems,
like partitioning or coloring problems.
Examples of solving linear programming problems with symmetries can be found in \cite{Margot}.
A review of results and techniques for solving a symmetric constraint programming problems can be found in \cite{Gent}.
Existence of symmetries in a given optimization problem facilitates
searching for a solution of it.
Usually such problem splits into $|G|$ identical parts,
where $|G|$ is an order of a symmetry group,
and it is enough to solve the problem only for an one part.
Solving given problem on a reduced part is called a 'symmetry breaking' procedure, because the reduced
problem looses its symmetry, \cite{Margot}, \cite{Gent}.
In this paper we utilize the technique of a 'symmetry breaking' to simplify solution of the
$G$-invariant maximum lifetime problem in sensor networks.
A 'symmetry breaking' of $G$-invariant problem (\ref{MinimaxDefMatrix}) is performed by
reduction of its solution to the optimal fundamental region $F_0 \subset S_{N}^{K}$ of the symmetry group $G$.
Because a selection of a fundamental region $F$ for given symmetry group $G$ is not unique,
and the problem cannot be reduced for every fundamental region,
we show in the paper how to construct the optimal one and we investigate its properties.
It is not always evident that for a given optimization problem with a symmetry group $G$
there exists a $G$-invariant solution of it.
The main result of this paper is a theorem, which states that for
a considered $G$-invariant maximum lifetime problem
there exists a symmetric ($G$-invariant) solution.
In the paper we investigate in detail properties of the isometry invariant solutions
of the problem in two dimensional sensor networks.
It seems, that the presented paper is a first attempt
of analyzing symmetries and utilize methods of solving optimization problems
in sensor networks by means of their symmetry groups.
\section{A space symmetry of the maximum network lifetime problem}
We consider a group $G$ of a one-to-one transformations of the two dimensional plane $R^2$ onto itself.
Elements $g$ of the group $G$ transform points $p$ of $R^2$ to some other points $p'=g(p)$ in $R^2$.
Because $S_{N}^{K} \subset R^2$, the points of the network $p_i$ are transformed under $g$ according
to the equation $p_i^{'}=g(p_i)$.
The transformed network $g(S_{N}^{K})$ by an element $g$ of $G$ we denote by $S_{N}^{'K}$. 
We assume, that
the functions $E_{i,j}(\bar{p})$ and $Q_{i}(\bar{p})$ in (\ref{MinimaxDefMatrix})
are invariant under transformations of the group $G$,
which means that $\forall g\in G$ we have
\begin{equation} \label{InvQp}
\left\{   \begin{array}{l}
E_{i,j}(g(\bar{p^{}}))= E_{i,j}(\bar{p}), \\
Q_{i}(g(\bar{p^{}}))=Q_{i}(\bar{p}),
\end{array} \right.
\end{equation}
where $g(\bar{p})=(g(p_1),....,g(p_{K+N}))$ and $p_i\in R^2$, $i\in [1,K+N]$. \\\\
{\bf Proposition 1}.
Let $G_{{\rm}}$ be a transformation group of two dimensional plane $R^2$,
$E_{i,j}(\bar{p})$ and $Q^{}_{i}(\bar{p})$ be $G$-invariant functions in $R^2$,
then the solution $q(\bar{p})$ of the maximum lifetime problem (\ref{MinimaxDefMatrix})
is $G$-invariant.\\
{\it Proof}.
The invariance of $q_{i,j}(\bar{p})$ under transformation group $G$
means that $\forall \; g\in G$, the equations
$q_{i,j}(g(\bar{p}))=q_{i,j}(\bar{p})$ are satisfied.
Since the functions $Q_i(\bar{p})$ are $G$-invariant,
the feasible set given by the second equation in (\ref{MinimaxDefMatrix}) is $G$-invariant
$$(q^{'} - q^{'T}) v^{}_{0} = Q(g(\bar{p}))=Q(\bar{p}).$$
The invariance of the feasible set under $G$ means that the scopes of the parameters $q$ and $q'$
in  (\ref{MinimaxDefMatrix}) for both networks $S^{K}_{N}$ and $S^{'K}_{N}$ are the same.
From the condition (\ref{InvQp}) and the invariance of a feasible set
it follows that the function $f(q,\bar{p})$, given by (\ref{ObjFunction}), is $G$-invariant
with respect to the $\bar{p}$ variable
$$\forall g\in G \;\;f(q,\bar{p})=f(q,g(\bar{p})).$$
From the above it follows that $f(q,\bar{p})$ is a constant function for any $g\in G$
and the solution $q_{i,j}(\bar{p})$ of (\ref{MinimaxDefMatrix}),
as a minimal value of $f(q,\bar{p})$, satisfies
$$\forall g\in G \;\;q_{i,j}(g(\bar{p}))=q_{i,j}(\bar{p}),$$
which is a $G$-invariance condition for $q(\bar{p})$. $\diamond$

As an example of the application of the above proposition, let us consider a data transmission cost energy matrix $E_{i,j}(\bar{p})$
to be a function of the Euclidean distance
\begin{equation} \label{EuclidMetric}
    d(p^{}_{i}, p^{}_{j})= \sqrt{(p^{1}_{i}-p^{1}_{j})^2 + (p^{2}_{i}-p^{2}_{j})^2 },
\end{equation}
between two elements $p^{}_{i}, p^{}_{j}$ of the network $S_{N}^{K}$, i.e.,
\begin{equation} \label{EijFunctionOfMetric}
E_{i,j}(\bar{p})=E_{i,j}(d(p^{}_{i}, p^{}_{j})).
\end{equation}
The group which leaves invariant the metric (\ref{EuclidMetric})
is a direct sum of two continuous abelian groups
$G = O_2 \oplus T$,
the orthogonal group $O_2$ and the translation group $T$ in $R^2$, \cite{Martin}.
If $Q(\bar{p})$ is invariant under $O_2 \oplus T$ and $E_{i,j}(\bar{p})$ is of the form (\ref{EijFunctionOfMetric}),
then we know from the Proposition 1 that the solution of (\ref{MinimaxDefMatrix})
is also $G$-invariant, which means that it must be a function of the distance $d$
and the $G$-invariant functions $Q(\bar{p})$
$$q_{i,j}(\bar{p}) = q_{i,j}(d,Q(\bar{p})).$$
Further analysis of the equations (\ref{MinimaxDefMatrix})
allow us to deduce that the solution of (\ref{MinimaxDefMatrix})
must be a linear function in $Q_i(\bar{p})$.
In the above example we see, that simple analysis of symmetries of the functions $E(\bar{p})$ and $Q(\bar{p})$
and the structure of the equations (\ref{MinimaxDefMatrix})
allows us to predict general form of its solution which greatly facilitates searching for it.
\section{An internal symmetry of the sensor network}
As an internal symmetry group $G$ of a sensor network $S_{N}^{K}$ we consider
a one-to-one transformations of $S_{N}^{K}$ onto itself.
Such groups are subgroups of a symmetric group $\Pi(S_{N}^{K})$.
Because we cannot mix the sensors and the data collectors,
the group $G$ must be a direct sum of two subgroups $\Pi(S_{N})$ and $\Pi(C_{K})$,
which transform sensors into sensors and data collectors into data collectors, see (\ref{SymmetryGroupForSN}).
For an action of $g\in G$ on elements $p_i$  of the network $S_{N}^{K}$
$$g(p^{}_{i})=p^{}_{g(i)}=p^{}_{i'},$$
we have the following transformation rules for $E$ and $Q$
$$\left\{   \begin{array}{l}
g(E_{i,j})= E_{g(i),g(i)} =E_{i',j'},\\
g(Q_{i})=Q_{g(i)}=Q_{i'}.
\end{array} \right.$$
In a matrix representation of $G$,
the elements $g\in G$ act on $S_{N}^{K}$ as a linear transformations
$p^{}_{i'}=\sum_{j}^{} g_{i',j}^{} \; p^{}_{j}$, which induces the following transformations
on the matrix $E$ and the vector $Q$
\begin{equation} \label{TransformForEAndQ}
E^{'}=gEg^{-1},\; Q^{'}=gQ.
\end{equation}
The next proposition shows that the problem (\ref{MinimaxDefMatrix})
is covariant under action of the group $\Pi(C_{K}) \oplus \Pi(S_{N})$.
It means, that for any transformation (\ref{TransformForEAndQ}) of $E_{i,j}$ and $Q_{i}$,
by an element $g\in \Pi(C_{K}) \oplus \Pi(S_{N})$,
the solutions $q$ and $q'$ of (\ref{MinimaxDefMatrix}) are related by the transformation $g$.\\\\
{\bf Proposition 2}.
Let $q_{}$ and $q'$ be solutions of the maximum lifetime problem (\ref{MinimaxDefMatrix})
for $(S_{N}^{K}, E, Q)$ and $(S_{N}^{'K}, E', Q')$ sensor networks,
where $S_{N}^{'K}=g(S_{N}^{K})$ and $g \in \Pi(C_{K}) \oplus \Pi(S_{N})$, then the diagram
\begin{equation} \label{TransfDiagram}  \begin{array}{l}
     (E,Q) \;\;\; \longrightarrow \;\;q \\
\;\;\;\;\downarrow^{g} \;\;\;\;\;\;\;\;\;\;\;\;\;\;\;\uparrow^{g^{-1}}\\
   (E', Q') \longrightarrow \;\;q'\\
\end{array}
\end{equation}
is commutative.

{\it Proof}. 
Since any permutation can be written as a product of transpositions,
it is enough to check (\ref{TransfDiagram}) for the transpositions
\begin{equation} \label{Transposition}  \begin{array}{l}
\forall_{i\neq r,r'} \; g(p_{i})=p_{i}, \;g(p_{r})=p_{r'}, \; p_r, p_{r'}\in S_N \;\;{\rm or }\;\; p_r, p_{r'} \in C_K.
\end{array}
\end{equation}
The feasible set, defined by the second equation in (\ref{MinimaxDefMatrix})
$$h_{i}(q,\bar{p})=0, \;\;\;i\in [K+1,K+N],$$
where
$h_{i}(q,\bar{p}) = \sum_{j=1}^{K+N}(q_{i,j} - q_{j,i}) - Q_{i}$,
is unchanged under the transpositions (\ref{Transposition}), because of the relations
$$g(h_r)=h_{r'},\;\;g(h_i)=h_{i}, \; i\neq r,r'.$$
The feasible set $h_{i}(q,\bar{p})=0$ for $i\in [1,K]$ is trivially $G$-invariant
because of the identity $Q_i=q_i^{(c)}$.
The transpositions (\ref{Transposition}) exchange the energies consumed by the $r$-th and $r'$-th sensors
$g(E_r)=E_{r'}$ and
other sensor energy levels remain unchanged
$g(E_{i})=E_{i}$, for $i\neq r,r'$, where
$E_{i}$ is given by (\ref{SensorEnergy}).
From this follows that the objective function of the maximum lifetime problem
given by (\ref{ObjFunction}) is invariant under (\ref{Transposition}).
The invariance of $f(q,E,Q)$ under $\Pi(C_{K}) \oplus \Pi(S_{N})$ means that $\forall g\in G$
and for $E'=g(E),Q'= g(Q), q'=g(q)$
the functions are equal, $f(q,E,Q) = f(q', E',Q')$,
and have the same minimal value with respect to the $q$ variable
$$\min_q f(q,E,Q) = \min_{q'} f(q', E',Q').$$
From the above equation it follows that, if $q_{i',j'}$ is a solution of (\ref{MinimaxDefMatrix})
for $E^{'}$, $Q^{'}$ then the matrix
$$q_{i,j}= g^{-1}(q_{i',j'})$$
is a solution of (\ref{MinimaxDefMatrix}) for $E^{}$, $Q^{}$. $\diamond$

In the next section, based on the result of the Proposition 2 we show that for
a $G$-invariant functions $E_{i,j}$ and  $Q_{i}$, where
$G$ is a subgroup of $\Pi(C_{K}) \oplus \Pi(S_{N})$,
there exists a $G$-invariant solution of (\ref{MinimaxDefMatrix}).
\subsection{The $G$ invariant solution of the maximum network lifetime problem }
The $G$-invariance of $E_{i,j}$ and  $Q_{i}$ means that for any transformation
(\ref{TransformForEAndQ}) we have
$$\left\{   \begin{array}{l}
E_{g(i),g(j)} = E_{i,j},\\
Q_{g(i)} = Q_{i}.
\end{array} \right.$$
In the following theorem we construct a $G$-invariant solutions of (\ref{MinimaxDefMatrix}).\\\\
{\bf Theorem 1}.
Let $G$ be a symmetry group of the set $S_{N}^{K}$,
which transforms sensors into sensors and data collectors into data collectors,
$E$ and $Q$ be a $G$-invariant functions in (\ref{MinimaxDefMatrix}),
then there exists a $G$-invariant solution $q_{}$ of (\ref{MinimaxDefMatrix}), i.e.,
\begin{equation} \label{gqg1}
\forall_{g\in G}\;\; g q_{} g^{-1} = q_{}.
\end{equation}
{\it Proof}.
We assume that for a given $G$-invariant matrix $E$ and a vector $Q$
there exists, not necessary a $G$-invariant, solution of (\ref{MinimaxDefMatrix}).
We denote it by $q_{0}$.
From the Proposition 2 we know that
$$ \forall g_m\in G \;\;\; q_m = g_m q_{0} g_m^{-1}$$
are solutions of (\ref{MinimaxDefMatrix}) for the same $E$ and $Q$.
We show that any linear combination of $q_m$
\begin{equation} \label{qlambda}
 q(\lambda_0,...,\lambda_{M-1})= \frac{1}{\lambda_0 + ...+\lambda_{M-1}}\sum_{m=0}^{M-1}\lambda_m q_m,
\end{equation}
where $\lambda_m\geq 0$, $\sum_m \lambda_m >0$, $M=|G_{{\rm}}|$
is also a solution of (\ref{MinimaxDefMatrix}).
We order the energies consumed by all sensors from $S_N$ in a decreasing sequence
\begin{equation} \label{Eqlambda}
 \forall m\;\;\; E_{1}(q_m)\geq ... \geq E_{N}(q_m),
\end{equation}
where $E_{i}^{}(q_m)$ is given by (\ref{SensorEnergy}) and $m\in [0,M-1]$.
For $\lambda_m\geq 0$, $\sum_m \lambda_m >0$, from (\ref{Eqlambda}) it follows that
\begin{equation} \label{EnOrdered}
\sum_{m} \lambda_m E_{1}(q_{m})\geq ...  \geq \sum_{m} \lambda_m E_{N}(q_{m}).
 \end{equation}
Because all $q_m$ are solutions of (\ref{MinimaxDefMatrix}) for the same functions $E, Q$,
then $\forall_m \; E_{1}(q_m)=E_{1}(q_0)$,
and (\ref{EnOrdered}) is equivalent to
$$(\sum_{m} \lambda_m) E_{1}(q_{0}) \geq  ...  \geq \sum_{m} \lambda_m E_{N}(q_{m}).$$
From the relation
$(\sum_{m} \lambda_m) E_{1}(q_{0}) = E_{1}(\sum_{m} \lambda_m q_{m})$
it follows that any matrix of the form (\ref{qlambda}) is a solution of (\ref{MinimaxDefMatrix}).
Now, it is easy to show that for a symmetry group $G$ of the problem (\ref{MinimaxDefMatrix})
the matrix
\begin{equation} \label{G-Invariant-qij}
 q = \frac{1}{|G|} \sum_{ g_m \in G} g_m q_{0} g_m^{-1}
\end{equation}
satisfies (\ref{gqg1}) and from (\ref{qlambda})-(\ref{EnOrdered}) it follows that
it is a solution of (\ref{MinimaxDefMatrix}).
The $G$-invariance of (\ref{G-Invariant-qij}) follows from the identity,
$\forall g_n\in G_{{\rm }}$
$$g_n q g_n^{-1} =\frac{1}{|G|}\sum_{g_m\in G} (g_n g_m) q_{0} (g_n g_m)^{-1}
= \frac{1}{|G|}\sum_{g_{m'}\in G}^{} g_{m'} q_{0} g_{m'}^{-1} = q,$$
where we used the fact that $\forall_{g_n\in G}$ $g_n G = G$.  $\diamond$
\section{Reduction of the maximum network lifetime problem to the subset of $S_{N}^K$ }
A $G$-invariant solution of the maximum network lifetime problem (\ref{MinimaxDefMatrix})
satisfy constraint (\ref{gqg1}),
which can be used to reduce the number of variables $q_{i,j}$
and thus simplify the solution of (\ref{MinimaxDefMatrix}).
On other hand, we would like to relate the shape of the network $S_{N}^K$
and its symmetry with the invariance of a solution $q_{i,j}$ of (\ref{MinimaxDefMatrix}).
This would allow us to determine a solution of (\ref{MinimaxDefMatrix})
based on the shape of the network $S_{N}^K$.
One can easily relate the shape of the network $S_{N}^{K}$ with
a symmetry of the data transmission cost energy matrix $E_{i,j}(\bar{p})$ by requiring
that the matrix elements $E_{i,j}(\bar{p})$ are functions of the distance $d(p_{i},p_{j})$ between
points in $S_{N}^{K}$.
Transformations which preserve the distance (\ref{EuclidMetric})
form a group of isometries of an Euclidean plane $R^2$.
Every isometry of a real Euclidean space is a composition of a translation and
an orthogonal transformation, \cite{Martin}.
For a finite set $S_{N}^{K}$ in $R^2$ every isometry
is an orthogonal transformation $g\in O_2$,
since there is no translation which transforms a finite set onto itself.
In this section we consider as a symmetry group $G$ of the sensor network $S_{N}^{K}$
subgroups of the orthogonal group $O_2$ and
assume that the functions $E_{i,j}(\bar{p})$ and $Q_i(\bar{p})$ are $O_2$-invariant.
The data transmission cost energy matrices $E(\bar{p})$, which are functions
of the distance between elements of the sensor network
are widely used in data transmission models in sensor networks.
For example, in \cite{Giridhar}
a maximum network lifetime problem was considered for a one dimensional,
regular sensor network with the data transmission cost energy matrix
of the form $E_{i,j} \sim |i-j|^{\alpha} e^{\gamma |i-j|}$, $\alpha \geq 2$, $\gamma\geq 0$.

One may ask, what is the general form of the $O_2$-invariant data transmission cost energy matrix $E_{}$.
The matrix $E=\sum_{n} \lambda_n d^{a_n}$,
which is a linear combination of powers of the matrix $d$, see (\ref{EuclidMetric}),
is trivially $O_2$-invariant.
The next simple lemma shows that $O_2$-invariant $E_{i,j}$ matrix can be build of powers of
the matrix elements of the metric $d_{i,j}$.\\\\
{\bf Lemma 1}. For the group $O_2$ of the orthogonal transformations of the Euclidean plane $R^2$
the matrix
\begin{equation} \label{Eijdija}
E_{i,j} = d(p^{}_{i}, p^{}_{j})^a, \;\; a\in R
\end{equation}
is $O_2$-invariant.\\
{\it Proof}.
Since any orthogonal transformation $g\in O_2$ act as permutation on the set $S_{N}^{K}$ and
$\forall g\in G$ $d(p^{}_{g(i)}, p^{}_{g(j)})=d(p^{}_{i}, p^{}_{j})$,
then for any $a\in R$ we have the following implication
$$d(p^{}_{i'}, p^{}_{j'})=d(p^{}_{i}, p^{}_{j}) \Rightarrow d(p^{}_{i'}, p^{}_{j'})^a = d(p^{}_{i}, p^{}_{j})^a,$$
from which follows the $O_2$-invariance of $E$ given by (\ref{Eijdija}).  $\diamond$

A linear action of the orthogonal group $O_2$ on the set $S_{N}^{K}$ and the results of the Lemma 1
allows us to write a general form of the $O_2$-invariant data transmission cost energy matrix $E_{i,j}$
\begin{equation} \label{GeneralFormOfE}
E_{i,j}(\bar{\lambda}, \bar{a})=\sum_{n}\lambda_{n} d_{i,j}^{a_n},
\end{equation}
where $\bar{\lambda}=(\lambda_1,...)$, $\bar{a}=(a_1,...)$ and $\lambda_n, a_n \in R$.
Most of the data transmission models in sensor networks
utilize the data transmission cost energy functions given by (\ref{GeneralFormOfE}).

Because a finite subgroup of the orthogonal group $O_{2}$ is
either a dihedral group $D_{M}$ or a rotation group $R^{(M)}$, \cite{Martin}, \cite{Grove},
we consider these groups as symmetries of the sensor network $S_{N}^{K}$
and the functions $E_{i,j}$ and $Q_{i}$.
We show, that in both cases the problem (\ref{MinimaxDefMatrix}) can be reduced to
subset of $S_{N}^{K}$ which we call an optimal fundamental region $F_0^{*} \subset S_{N}^{K}$.
For the dihedral symmetry group $D_{M}$ the optimal fundamental region can be easily determined.
In case of the rotation group $R^{(M)}$ the optimal fundamental region must be determined
for a particular distribution of elements of $S_{N}^{K}$ over the plane $R^2$.
Existence of a reduction of the maximum network lifetime problem (\ref{MinimaxDefMatrix}) to the
optimal fundamental region will be proven under two assumptions.
We assume, that two sensors cannot exchange the data
\begin{equation} \label{NoLoopEq}
 q_{i,j}\neq 0 \; \Rightarrow \; q_{j,i}= 0, \;\; p_i, p_j \in S_N,
\end{equation}
which means that if the sensor $p_i$ sends some data to the sensor $p_j$,
then the sensor $p_j$ cannot send any data to $p_i$.
Second assumption is that we consider only a $G$-invariant $E_{i,j}$ matrices having the property
\begin{equation} \label{dijLEQEij}
d(p_{i},p_{j})\leq d(p_{i'},p_{j'}) \Rightarrow E_{i,j}\leq E_{i',j'},
\end{equation}
which means that for such matrices the cost of data transmission grows whenever the distance between elements
of the network $S_{N}^{K}$ grows.
\subsection{The sensor network with a dihedral symmetry group}
The dihedral group $D_{M}$ is a symmetry group of a regular polygon with $M$ sides.
It is a semidirect product $R^{(M)} \rtimes S^{(M)}$ of the rotation group $R^{(M)}$ and the reflection group $S^{(M)}$.
It consists of $2M$ elements, $M$ reflections $\{S_m\}_{m=0}^{M-1}$ and $M$ rotations $\{R_m\}_{m=0}^{M-1}$.
Because any rotation can be represented as a product of even reflections then the dihedral group
can be generated by $M$ reflections, $D_{M}=\langle S_m \rangle_{m=0}^{M-1}$.
For a given set $S_{N}^{K}$ with the symmetry group $D_{M}$,
a subset $F$ of $S_{N}^{K}$ is called a fundamental region for $D_{M}$,
if $\forall_{ g\neq I \in D_{M}}$ $F\cap g(F)=\emptyset$ and
$S_{N}^{K} = \bigcup_{g\in D_{M}} g(F)\cap S_{N}^{K}$, \cite{Grove}.
The first requirement means that there are no elements of the network $S_{N}^{K}$ on the reflection lines $X_m$.
The second requirements states that the set $S_{N}^{K}$ is a disjoint union of $|D_{M}|=2M$
subsets $F_m$, and
$S_{N}^{K}=\bigcup_{m\in [0,2M-1]} F_{m}$.
The requirement, that there are no elements of the network $S_{N}^{K}$ on the reflection
lines $X_m$, $m\in [0,M-1]$, can be written as
$$ \forall_{p\in S_N^K} \;\;{\rm St}(p,D_{M})=\{ g\in D_{M} \;:\; g(p)=p\}=\{I\},$$
which means that the stabilizer of any point $p\in S_N^{K}$ is trivial.
A fundamental region $F$ of the $S_{N}^{K}$ set can be selected in many ways.
Among fundamental regions in $S_{N}^{K}$ there is only one for which the $D_{M}$-invariant problem
(\ref{MinimaxDefMatrix}) can be reduced.
Let $V_0$ be a region between $X$-axis, $X\geq 0$, and the reflection line $X_1\geq 0$.
We assume, that the element $S_0$ of the dihedral group $D_{M}$ is a reflection along the $X$-axis and
$S_1$ along the $X_1$ line.
There exists only one fundamental region $F_{0}$ which is a subset of $V_0$.
From the set $F_{0}$, by the following sequence of transformations
$F_{m} = S_m(F_{m-1})$ for $m\in [1,M-1]$ and
$F_{M+m} = S^{-1}_m(F_{M+m-1})$ for $m\in [0,M-1]$,
where $S_m$ are reflections of $D_{M}$,
we can generate $|G|=2M$ disjoint sets $F_{m}$ and represent the network $S_{N}^{K}$ as a sum of them,
$S_{N}^{K} =\bigcup_{m=0}^{2M-1}F_{m}$.
We show that the fundamental region $F_{0}$ is an optimal one,
which means that there exists a $D_{M}$-invariant solution $q_{i,j}$ of (\ref{MinimaxDefMatrix})
for which $q_{i,j}\neq 0 \Rightarrow p_i, p_j \in F_{m}$ and
the matrix $q_{i,j}$ can be factorized into $2M$ identical sub-matrices, a one sub-matrix for each region $F_{m}$.
In other words, we show that
the regions $F_{m}$ are closed for data transmission
and inside each of them the data transmission paths are identical,
so it is enough to find the solution of (\ref{MinimaxDefMatrix}) in one of them,
for example in $F_{0}$.
The reduction of the $D_{M}$-invariant problem (\ref{MinimaxDefMatrix}) to the subset of $S_{N}^{K}$
is also possible when the data collectors are located on the reflection lines $X_m$.
Let us denote by $\partial V_{0}$ the border of the angle $V_0$.
The set $\partial V_{0}$ is a sum of two half-lines $X_0 \geq 0$ and $X_1\geq 0$.
By $C_{0}$ and $C_{1}$ we denote the set of data collectors which are located
on the half-line $X_0 \geq 0$ and $X_1\geq 0$ respectively.
The set of data collectors $C^{(X)}$ which lie on all the reflections lines is the sum of
the sets $S_{m}(C_{0})$ and $S_{m}(C_{1})$, i.e.,
$C^{(X)} = \bigcup_{m=0}^{M-1} (S_{m}(C_{0}) \cup S_{m}(C_{1})),$
and it is a $D_{M}$-invariant.
We add the elements of the set $C^{(X)}$ to the sensor network $S_{N}^{K}$
and obtain a $D_{M}$-invariant set $S_N^K \cup C^{(X)}$.
An example of the set $F_{0} \cup C_{0} \cup C_{1}$ for $D_{4}$ group is shown in the Figure 1.
In the fundamental region $F_{0}$ there are seven sensors and one data collector,
three data collectors are located at the border of $V_{0}$ and lie in the sets $C_{0}$ and $C_{1}$.

\begin{figure}[!ht]
\begin{center}
\includegraphics{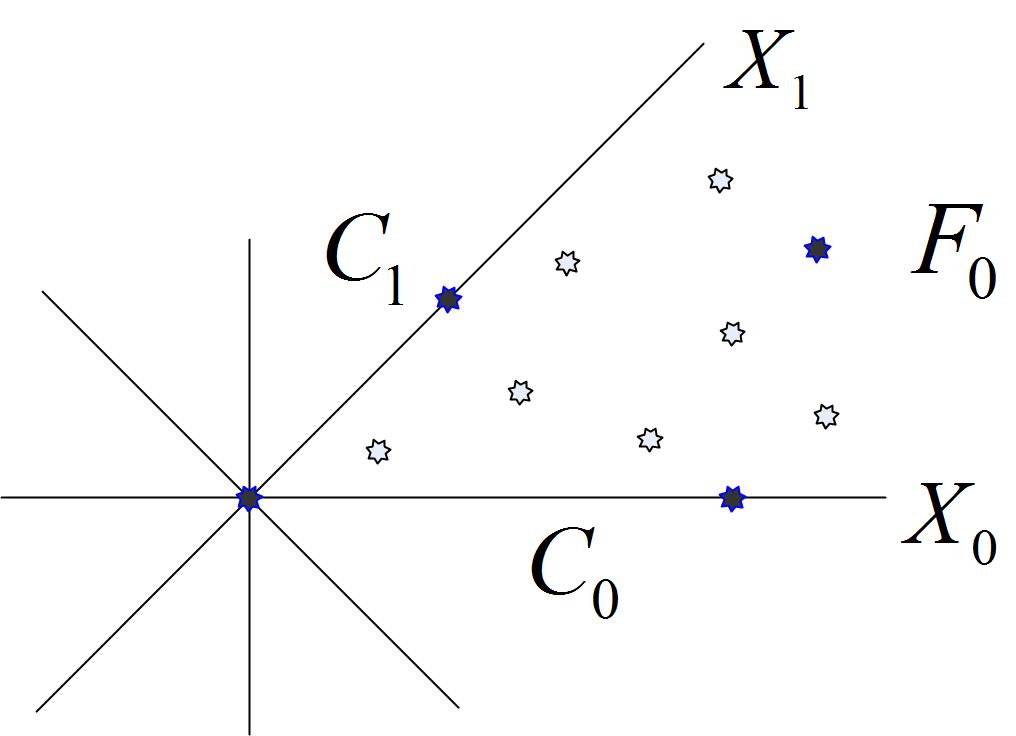}
\caption{Fundamental region $F_{0}^{}$ for a dihedral group $D_4$ and nonempty sets $C_{0}$ and $C_{1}$.}
\label{Fig03FzeroDih}
\end{center}
\end{figure}
%

The following proposition shows that the $D_{M}$-invariant problem (\ref{MinimaxDefMatrix})
over $D_{M}$-invariant network $S_N^K \cup C^{(X)}$,
with the data transmission cost energy matrix $E_{i,j}$ having the property (\ref{dijLEQEij}),
can be reduced to the
region $F_{0} \cup C_{0} \cup C_{1}$.\\\\
{\bf Proposition 3}.
Let
$D_{M}$ be a symmetry group of the set $S_N^K \cup C^{(X)}$ with trivial stabilizer for each $p_i\in S_N^K$,
$F_{0} \subset V_0$ a fundamental region in $S_N^K$ for the group $D_{M}$ and
let $C_{0}$, $C_{1}$ are the sets of data collectors which lie on the border of $V_0$,
then the solution of a $D_{M}$-invariant problem (\ref{MinimaxDefMatrix}) with $E$ satisfying (\ref{dijLEQEij})
can be reduced to $F_{0} \cup C_{0} \cup C_{1}$.\\
{\it Proof}.
From the Theorem 1 we know that for a $D_{M}$-invariant problem (\ref{MinimaxDefMatrix})
there exists a $D_{M}$-invariant solution.
We show, that the $D_{M}$-invariant solution $q_{i,j}$ of (\ref{MinimaxDefMatrix})
can be factorized to $2M$ copies,
and each copy is identical to the solution of (\ref{MinimaxDefMatrix})
in the region $F_{0}\cup C_{0} \cup C_{1}$.
For a given reflection $S_m \in D_{M}$ we write the set $S_N^K \cup C^{(X)}$ as sum of three subsets
$S_N^K= (S_N^K)^{(1)}_{m} \cup C^m \cup (S_N^K)^{(2)}_{m}$,
where $C^m$ is a set of data collectors which lie on the reflection line $X_m$
and
$(S_N^K)^{(a)}_{m}$, $a=1,2$ are sets of sensor network $S_N^K$ elements which lie on both sides of $X_m$.
On the reflection line $X_m$ there is a set of data collectors isometric to
the set $C_{0} \cup C_{1}$ for $M$ odd,
$C_{0} \cup C_{0}$ for $M$ even and $m$ even and
$C_{1} \cup C_{1}$ for $M$ even and $m$ odd.
The elements of $(S_N^K)^{(a)}_{m}$ we denote by $p_{a.i}$, $a=1,2$, where $S_m(p_{1.i})=p_{2.i}$.
The invariance of the solution $q$ of (\ref{MinimaxDefMatrix})
under transformation $S_m q (S_m)^{-1}=q$ can be written in the form
\begin{equation} \label{ConstrRefl1}\left\{   \begin{array}{l}
q_{1.i,2.j}=q_{2.i,1.j}, \\
q_{1.i,1.j}=q_{2.i,2.j},\\
q_{1.i,2.i}=q_{2.i,1.i}=0,\\
q_{1.j,2.j}=q_{2.j,1.j}=0,
\end{array} \right.
\end{equation}
for all $p_{a.i}\in (S_N^K)^{(a)}_{m}$, $a=1,2$.
The second equation in (\ref{ConstrRefl1}) follows from the requirement (\ref{NoLoopEq}).
From the geometric properties of the reflection symmetry we have $d(p_{1.i},p_{1.j}) \leq d(p_{1.i},p_{2.j})$,
and because of the assumption (\ref{dijLEQEij}), we get the set of inequalities
$E_{1.i,1.j}\leq E_{1.i,2.j}$ and $E_{2.i,2.j}\leq E_{2.i,1.j}$.
From these inequalities it follows that for a $D_{M}$-invariant solution $q_{}$ of (\ref{MinimaxDefMatrix})
for which $q_{1.i,2.j}=q_{2.i,1.j}\neq 0$,
we can find a solution $q'$ for which $q'_{1.i,2.j}=q'_{1.j,2.i}=0$
and $q'_{1.i,1.j}=q'_{2.i,2.j}=q_{1.i,2.j}$.
This means that we can construct a $D_{M}$-invariant solution $q'$ of (\ref{MinimaxDefMatrix}) for which
there is no data transmission across the reflection line $X_m$ and
inside the sets $(S_N^K)^{(1)}_{m}$, $(S_N^K)^{(2)}_{m}$ the data transmission is given by the same solution.
From the definition of the maximum network lifetime problem (\ref{MinimaxDefMatrix}) and the property
(\ref{dijLEQEij}) of the data transmission cost energy matrix $E$
it follows that the sensors always send their data to the nearest data collector
$$\forall \; p_n \in S_N, p_k\in C_K\cup C^{(X)}, \;\; q_{n,k}\neq 0 \Rightarrow  \forall p_{k'}\in C_{K}^{}\cup C^{(X)} \;\; d(p_n, p_k) \leq d(p_n, p_{k'}).$$
This means that the data collectors cannot receive any data from senors which lie behind the reflection line $X_m$.
These properties are valid for any reflection $S_m$, $m\in [0,M-1]$ of $D_{M}$ and
from this it follows that the data is not sent across any reflection line $X_m$, $m\in [0,M-1]$.
Because of the symmetry, inside each of the $2M$ regions the solutions of (\ref{MinimaxDefMatrix})
are identical and can be represented by a solution in $F_{0} \cup C_{0} \cup C_{1}$.  $\diamond$

\begin{figure} [!ht]
\begin{center}
\includegraphics{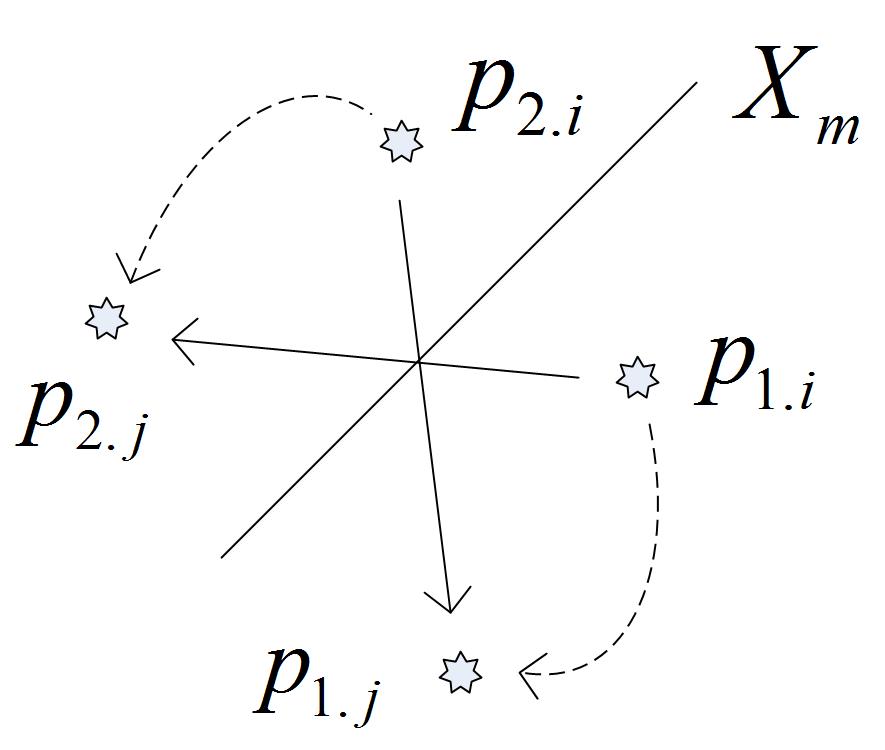}
\caption{The dashed arrows indicate the optimal data transmission path between two sensors.}
\label{Fig01Sm}
\end{center}
\end{figure}

In Figure \ref{Fig01Sm}, the dashed arrows indicate the optimal data transmission path
between sensors which lie on both sides of the reflection line $X_m$.

In the Proposition 3 the assumption that the set $C^{(X)}$ is $D_{M}$-invariant can be omitted.
In such case the network $S_N^K \cup C^{(X)}$ and the solution of the problem (\ref{MinimaxDefMatrix})
can be factorized into $2M$ parts,
but the solutions in each part are different,
due to the difference of the data collector sets on various reflection lines $X_m$.
\subsection{The sensor network with a rotation symmetry group}
A rotation group in $R^2$ is a cyclic group
generated by $M$ elements
$R^{(M)}=\langle R_{m} \rangle_{m=0}^{M-1}$,
where $R_{m}$ denotes a rotation by the angle $\alpha_m=\frac{2\pi}{M}m$.
We assume that the rotations are around the point $p_0=(0,0)\in R^2$.
The point $p_0$ is unique for which the stabilizer is non-trivial
and it is equal to the whole group $R^{(M)}$, ${\rm St}(p_0,R^{(M)})=R^{(M)}$.
From this reason, we assume that at the point $p_0$ there is no element of the network $S_{N}^{K}$.
If it is necessary to consider a sensor network with an element located at $p_0$,
then we will build a $R^{(M)}$-invariant sensor network $S_{N}^{K}\cup C^{(0)}$,
where $C^{(0)}$ is a set which consists of a one element located at $p_0$, a data collector.
Let us denote by $V_0$ the area between X-axis, $X\geq 0$, and the half-line $p^2=\tan[\alpha_1] p^1$,
$p^1\geq 0$, where $\alpha_{1}=\frac{2\pi}{M}$ and $(p^1,p^2)\in R^2$.
For the set $S_{N}^{K}$ there exists only one fundamental region $F_{0}$ in $S_{N}^{K}$ which is a subset of $V_{0}$.
By rotation of $F_{0}$ by elements of $R^{(M)}$
$$R_m(F_{0})=F_{m},$$
we can obtain $M$ regions, $F_{m} \subset V_{m}$, $m\in [0,M-1]$, where
$V_{m}=R_m(V_{0})$, such that
$S_N^K = \bigcup_{m=0}^{M-1} F_{m}$
and
$\bigcap_{m=0}^{M-1} F_{m}=\{\emptyset \}$.
We describe the properties of a $R^{(M)}$-invariant solution of (\ref{MinimaxDefMatrix}) in
terms of orbits of the symmetry group $R^{(M)}$.
The orbit of the point $p \in S_{N}^K$ under action of the group $R^{(M)}$ is a subset of $S_{N}^K$
$${\rm Orb}(p,R^{(M)})=\{ p \in S_{N}^K : p = g(p), g \in R^{(M)} \}.$$
Since we assumed that $p_0 \notin S_{N}^{K}$, then
the fundamental region $F$ for $R^{(M)}$ can be defined as a set of orbits
$F = S_{N}^{K}/\sim_{R^{(M)}},$
where for $p_1 \neq p_2$, $p_1 \sim_{R^{(M)}} p_2$
$\Leftrightarrow$ $\exists g\in R^{(M)} : g(p_1) = p_2$.
The points of the set $F_{m}$ we denote by $p_{m.i}$,
where the number $m\in [0,M-1]$ indexes the elements of the $i$-th orbit.
We will count the points $p_{m.i}$ on the $i$-th orbit anticlockwise starting from the $X$-axis, $X\geq 0$.
The numbers $m\in [0,M-1]$ and $i$ uniquely identify the points of the set $S_N^K$.
For the set $S_{N}^{K}\cup C^{(0)}$, $i\in [0,\frac{N+K}{M}]$,
and for $S_{N}^{K}$, $i\in [1,\frac{N+K}{M}]$.
The effect of rotation of the point $p_{n.i}$ by the angle $\alpha_m=\frac{2\pi}{M}m$, $m\in [0,M-1]$
can be written by the formula
$$R_{m}(p_{n.i})=p_{(n+m).i},$$
where $n+m$ denotes $(n+m)|_{{\rm mod} M}$.
For $p_{0.0}\in C^{(0)}$, $R_{m}(p_{0.0})=p_{0.0}$.
The invariance of the solution of (\ref{MinimaxDefMatrix}) under $R_{k}$ rotation
can be rewritten in the form
\begin{equation} \label{RqR-1}
R_{k} q R_{k}^{-1} = q \;\;\Rightarrow \;\;
q_{(m+k).i,(n+k).j} = q_{m.i,n.j}.
\end{equation}
The following proposition states that for $R^{(M)}$-invariant maximum lifetime problem (\ref{MinimaxDefMatrix})
for the sensor network $S_{N}^{K}$
there exists a $R^{(M)}$-invariant solution
for which there is no data transmission between sensors which lie on the same orbit.\\\\
{\bf Proposition 4}.
Let $q$ be a solution of $R^{(M)}$-invariant problem (\ref{MinimaxDefMatrix}) for the $S_{N}^{K}$ network,
then the sensors from the same orbit do transmit data to each other,
i.e.,
$$\forall_{i,m,n} \; p_{m.i}, p_{n.i} \in S_N \;\; q_{m.i,n.i} = 0.$$
{\it Proof }.
From the Theorem 2 we know that for $R^{(M)}$-invariant maximum lifetime problem (\ref{MinimaxDefMatrix})
there exists a $R^{(M)}$-invariant solution $q$.
For such solution, if the sensor $p_{m.i}$ sends $q_{m.i, n.i}$ of data to the $p_{n.i}$ sensor then,
from (\ref{RqR-1}) we know that for $k=m-n$,
the $p_{(2m-n).i}$ sensor sends the same amount of data
$q_{(2m-n).i,m.i} = q_{m.i,n.i}$ to the $p_{m.i}$ sensor.
Because any amount of data which is sent by the $p_{m.i}$ sensor to the sensor on the same orbit 'returns' to it,
then we can find a $R^{(M)}$-invariant solution $q'$ of (\ref{MinimaxDefMatrix})
for which $\forall_{i,m,n} \; q'_{m.i,n.i}=0.$  $\diamond$

In the next proposition we prove that if the requirement (\ref{dijLEQEij})
for the data transmission cost energy matrix $E_{i,j}$ is satisfied,
then there exists a $R^{(M)}$-invariant solution $q$ of (\ref{MinimaxDefMatrix})
for which a sensor from one orbit sends its data to the nearest sensor from another orbit.\\\\
{\bf Proposition 5}.
Let $q$ be a solution of $R^{(M)}$-invariant problem (\ref{MinimaxDefMatrix})
for $S_{N}^{K}\cup C^{(0)}$ sensor network with $E_{i,j}$ satisfying (\ref{dijLEQEij}),
then for any sensor $p_{m.i}$ and any element $p_{n.j}$ of $S_{N}^{K}\cup C^{(0)}$
from different orbits, $i\neq j$,
the only non zero element of the matrix $q_{m.i, n.j}$ has the property
$$\forall_{i\neq j, m, n} \;\; q_{m.i, n.j} \neq 0  \;\Rightarrow  \;
\forall_{p_{n'.j}\in S_N} \;\; d(p_{m.i},p_{n.j}) \leq  d(p_{m.i},p_{n'.j}).$$
{\it Proof.}
Let us assume that for $R^{(M)}$-invariant solution $q$ of (\ref{MinimaxDefMatrix})
the $p_{m.i}$ sensor sends to the sensor or to the data collector $p_{n.j}$
the amount $q_{m.i,n.j}$ of data, $i\neq j$.
From (\ref{RqR-1}) we know that the same amount of data is sent from
the $p_{(m+k).i}$ sensor to the $p_{(n+k).j}$ sensor or to the data collector, $k\in [0,M-1]$.
As a result, each element of a sensor network from the $j$-th orbit receives the same amount
of data $q_{m.i, n.j}$ from a one sensor from the $i$-th orbit.
Due to the assumption that the data transmission cost energy matrix $E_{i,j}$ satisfies (\ref{dijLEQEij}),
the minimum energy of sending the amount of data $q_{m.i, n.j}$ by the $p_{m.i}$ sensor
from the $i$-th orbit to the element $p_{n.j} \in S_{N}^{K}\cup C^{(0)}$ from the $j$-th orbit
is achieved when the distance between $p_{m.i}$ and $p_{n.j}$ is minimal,
i.e., it has the property
$\forall_{p_{n'.j} \in S_{N}^{K} \cup C^{(0)}}$  $d(p_{m.i},p_{n.j}) \leq  d(p_{m.i},p_{n'.j})$.  $\diamond$

\begin{figure}[!ht]
\begin{center}
 \includegraphics{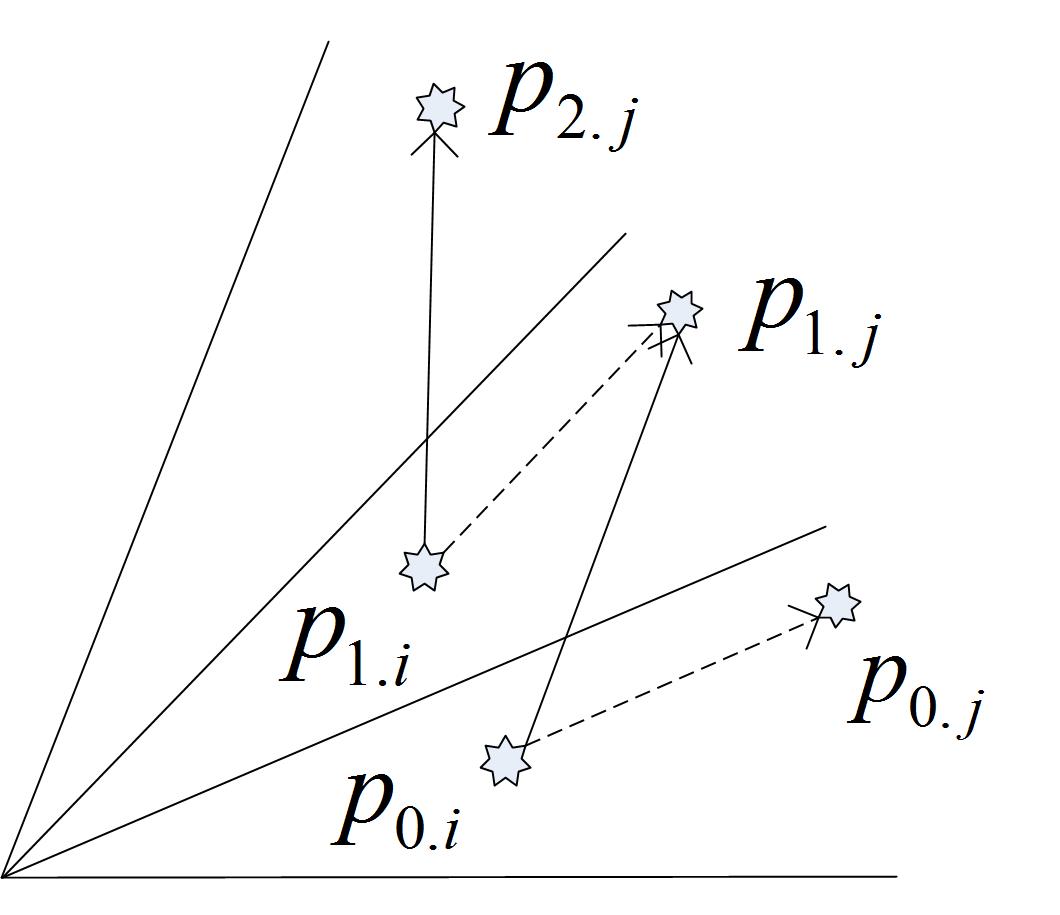}
\caption{The dashed arrows indicate the optimal data transmission path between elements of $S_{N}^{K}$ which lie on different orbits.}
\label{Fig04Rm}
\end{center}
\end{figure}

In Figure \ref{Fig04Rm} the dashed arrows indicate the optimal data transmission path between
sensors $p_{m,i}$ from the $i$-th orbit and sensors or data collectors $p_{m,j}$ from the $j$-th orbit, $m\in [0,M-1]$.

\begin{figure}[!ht]
\begin{center}
 \includegraphics{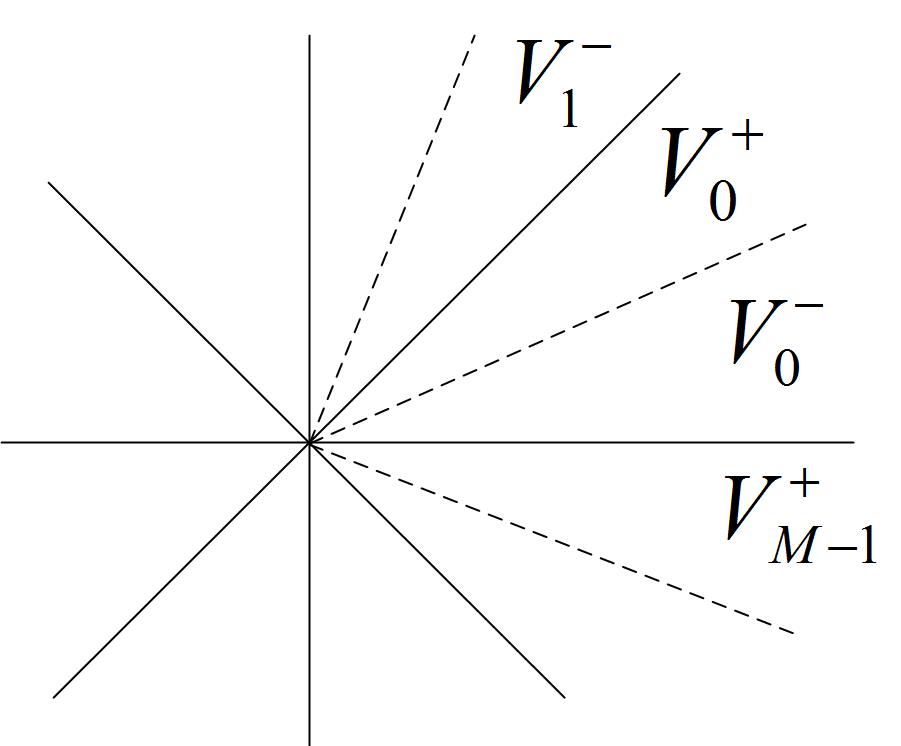}
\caption{Location of the regions $V_{0}^{\mp}$ in $V_{0}^{}$.}
\label{Fig02Vplus}
\end{center}
\end{figure}

Because the rotation group $R^{(M)}$ is abelian the constrains $R_m q R_m^{-1}=q$, $m\in [0,M-1]$
which satisfy the $R^{(M)}$-invariant solution of (\ref{MinimaxDefMatrix}) are not very restrictive.
In general, from the rotation invariance we cannot determine the optimal fundamental region $F_0^{*}$ in $S_{N}^{K}$,
i.e., a region to which the problem (\ref{MinimaxDefMatrix}) can be reduced.
The next three propositions describe the size and location of
the optimal fundamental region $F_0^{*}$ in $S_{N}^{K}$.
Let us denote by $V_0^{\pm}$ two sub-regions of $V_0$ such that
$$V_0 = V_0^{-}\cup V_0^{+},$$
where the points of $V_0^{-}$ lie on or between the $X$-axis, $X\geq 0$ and
the half-line $p^2=\tan[\frac{\alpha_1}{2}] p^1$, $p^1\geq 0$.
The region $V_0^{+}$ lie above of $V_0^{-}$.
The Figure \ref{Fig02Vplus} shows the location of the regions $V_{0}^{\mp}$ in $V_{0}^{}$.

The regions $V_m^{\pm}$ can be obtained by rotating $V_0^{\pm}$ by elements of
the group $R^{(M)}$, $V_m^{\pm}=R_m(V_{0}^{\pm})$.
The next proposition shows, that the sensors which lie in $V_{m}^{}$
can send their data to the data collectors or other sensors only when they lie
in $V_{m}^{}$ or in the neighboring regions $V_{m\mp 1}^{\pm}$.\\\\
{\bf Proposition 6}.
Let $q$ be a solution of $R^{(M)}$-invariant problem (\ref{MinimaxDefMatrix})
for $S_{N}^{K}\cup C^{(0)}$ network with $E_{i,j}$ satisfying (\ref{dijLEQEij}),
then the sensors $p_{m.i}$ from the region $V_{m}$
can send their data to the elements $p_{m'.i}$ of the sensor network $S_N^K \cup C^{(0)}$
only when they lie in $V_{m}^{} \cup V_{m \mp 1}^{\pm}$, i.e.,
$$
\forall_{i\neq j} \;\;
p_{m.i}\in V_{m}, \; q_{m.i,m'.j} \neq 0, \Rightarrow
p_{m'.j}\in V_{m}^{} \cup V_{m - 1}^{+}
\;\;
{\rm or}\;\;
p_{m'.j}\in V_{m}^{} \cup V_{m + 1}^{-}.$$
%
{\it Proof}.
For a sensor $p_{m.i}\in V_{m}^{+}$ from the $i$-th orbit which sends $q_{m.i,n.j}$ of data to
the sensor $p_{n.j}\in V_{n}$ from the $j$-th orbit, $i\neq j$, we can find a sensor $p_{m'.j}$ from
the $j$-th orbit which $p_{m'.j}\in V_{m}^{}\cup  V_{m+1}^{-},$
and $d(p_{m.i},p_{m'.j})=\min_{n'} d(p_{m.i},p_{n'.j})$.
Because of (\ref{dijLEQEij}) we can find $R^{(M)}$-invariant solutions $q$ of (\ref{MinimaxDefMatrix}),
for which the sensors from the set $V_{m}^{+}$ send their data
to the sensors or data collectors from $V_{m}^{}\cup V_{m-1}^{-}$.
Similarly, sensors $p_{m.i}$ from the set $V_{m}^{-}$ can send their data to the
elements of the sensor network from the subset $V_{m}^{}\cup V_{m-1}^{+} \subset S_{N}^{K}\cup C^{(0)}$
and the proposition is proven. $\diamond$

From the Proposition 5 we know that there exists
a $R^{(M)}$-invariant solutions $q$ of (\ref{MinimaxDefMatrix}) for which
a sensor from a given orbit sends its data only to a one sensor or a data collector from another orbit.
The following proposition describes conditions under which a fundamental region $F_0$ for $R^{(M)}$
is the optimal one. \\\\
{\bf Proposition 7}.
Let $F_0$ be a fundamental region in $S_{N}^{K}$
and the set $F_0\cup C^{(0)}$ fulfills the requirement
$$\forall p_{0.i}\in F_0\cup C^{(0)}, \forall_{j}\; d(p_{0.i},p_{0.j}) =\min_{m} d(p_{0.i},p_{m.j}),$$
$m\in [0,M-1]$,
then the solution of the $R^{(M)}$-invariant problem (\ref{MinimaxDefMatrix}) for $S_{N}^{K} \cup C^{(0)}$
with $E_{i,j}$ satisfying (\ref{dijLEQEij})
can be restricted to the set $F_0\cup C^{(0)}$.\\
{\it Proof}. From the Proposition 5 we know that there exists
$R^{(M)}$-invariant solutions $q$ of (\ref{MinimaxDefMatrix})
for which sensors send their data to the nearest sensor or data collector from other orbits.
We select the sensor $p_{0.1}\in S_{N}^{K}\cup C^{(0)}$ from the first orbit of $R^{(M)}$ and
build a set $F_{0} \cup C^{(0)}$ by picking up a one element $p_{0,j}$ from each orbit, such that
$$\forall_{i}\; d(p_{0.1},p_{0.i}) =\min_{m} d(p_{0.1},p_{m.i}).$$
If all elements $p_{0.i}$ of a constructed set $F_{0}\cup C^{(0)}$ have the property, that
from the inequality $\forall_{j}$ $d(p_{0.i},p_{m'.j}) =\min_{m} d(p_{0.i},p_{m.j})$ it follows
that $p_{m'.j}\in F_{0}\cup C^{(0)}$, then from the Proposition 5 we know that the set is closed for data transmission.
This means that for $p_{0.i}\in F_0, p_{m,j} \notin F_0 \Rightarrow q_{0.i,m.j}=0$.
Because $F_{0}$ is a fundamental region, then $S_{N}^{K}=\cup_{m=0}^{M-1}R_m (F_0)$,
and the solution of (\ref{MinimaxDefMatrix}) splits into $M$ copies, one for each region $R_m(F_0)\cup C^{(0)}$.  $\diamond$

The next proposition describes the location of the optimal fundamental region $F_0^{*}$
for $R^{(M)}$-invariant solutions of (\ref{MinimaxDefMatrix}) in the set $S_{N}^{K}$.\\\\
{\bf Proposition 8}.
Let $q$ be a solution of $R^{(M)}$-invariant problem (\ref{MinimaxDefMatrix}) for $S_{N}^{K}$ network
with $E_{i,j}$ satisfying (\ref{dijLEQEij}), then the optimal fundamental region $F_0^{*}$
is a subset of $V_{1}^{+} \cup V_0 \cup V_{M-1}^{-}$.\\
{\it Proof}. Follows from the Proposition 6. $\diamond$

It is easy to see that the Proposition 6 is also valid if we consider a sensor network with
a data collector located at the point $p_{0}$, i.e., for the $S_{N}^{K}\cup C^{(0)}$ network.
\section{Conclusions}
We have analyzed a continuous and discrete symmetries of the maximum lifetime problem in two dimensional
sensor networks $S_{N}^{K}$ built of $K$ data collectors and $N$ sensors.
We showed that, invariance of the problem under a continuous group of transformations $G$
implies that the solution is also $G$-invariant and can be expressed
in terms of the symmetry group invariants.
As we showed, this fact greatly facilitates searching for a strict or approximate solution of the problem.
In the paper, we also investigated properties of the solutions of the maximum lifetime problem
for sensor networks $S_{N}^{K}$ invariant under transformation groups $G$ which are subgroups
of the symmetric group $\Pi(C_{K}) \oplus \Pi(S_{N})$, where $C_{K}$ and $S_{N}$ are subsets of
$S_{N}^{K}$ which consist of the data collectors and sensors respectively.
We showed that for such groups a $G$-invariant maximum lifetime problem has a $G$-invariant solution.
In the paper we analyzed in detail invariance of the sensor network and solutions of the problem
under group of isometry transformations $O_2$ in $R^2$.
Constrains which follow from the $O_2$-invariance of a solution allowed us to
reduce it to a subset, an optimal fundamental region of the network.
The fact that, the $G$-invariant maximum lifetime problem and its solution
can be factorized and reduced to the fundamental region of the symmetry group $G$
can be utilized to design sensor networks with symmetries and with known solution
in the optimal fundamental region of the network.


\end{document}